\documentclass[review,12pt]{elsarticle}

\usepackage{showlabels}
\usepackage[colorlinks=true, pdfstartview=FitV, linkcolor=red, citecolor=blue, urlcolor=blue]{hyperref}
\usepackage{bbm}
\usepackage{graphicx}
\usepackage{amsmath}
\usepackage{amsfonts,amsbsy}
\usepackage{amssymb}

\newcommand{\beq}{\begin{equation}}
\newcommand{\eeq}{\end{equation}}
\newcommand{\bea}{\begin{eqnarray}}
\newcommand{\eea}{\end{eqnarray}}

\usepackage{mathtools}

\begin{document}

%\preprint{ }

\title{The MV Model of the Color Glass Condensate for a Finite Number of Sources Including Coulomb Interactions}
\author[bnl,ccnu]{Larry McLerran}
%\email{mclerran@bnl.gov}
\author[rbrc]{Vladimir V. Skokov}
\address[bnl]{
Department of Physics, Brookhaven National Laboratory, 
Upton, NY 11973}
\address[rbrc]{RIKEN/BNL, Brookhaven National Laboratory, 
Upton, NY 11973}
\address[ccnu]{Physics Dept, China Central Normal University, Wuhan, China}

%\email{vskokov@quark.phy.bnl.gov}

\begin{abstract}
We modify the McLerran-Venugopalan model to include only a finite number of sources of color charge.  We argue that Coulombic interactions between these color charges generates
a source-source correlation function that properly includes the effects of color charge screening,
a generalization of Debye screening for the Color Glass Condensate.
Such a model may be useful for computing angular harmonics of flow measured in high energy hadron collisions for small systems. 
In this paper we provide a basic formulation of the problem on a lattice. 
\end{abstract}

\maketitle

\section{Introduction}

The McLerran-Venugopalan (MV) model of the Color Glass Condensate  (CGC) has provided a useful phenomenology of high energy
hadronic processes~\cite{McLerran:1993ni,McLerran:1993ka}.  This model is however lacking in one essential aspect:  There 
is no color charge screening.
This lack of color screening follows from the form of the charge density correlation function
\begin{equation}
	\langle  \rho^a(y_1,\vec{x}_{1}) \rho^b(y_2, \vec{x}_{2}) \rangle = \frac{ d \mu^2 }{d y} \delta^{ab} \delta(y_1-y_2) \delta^{(2)}(\vec{x}_1-\vec{x}_2)\,.
\end{equation}
In this equation, we take the charge densities to be local both in rapidity  and in two dimensional transverse coordinate.  
The locality in rapidity is required for what we wish to do in the following analysis.  Upon integrating over coordinate
$y_1, \vec{x}_1$, we see that we get a non-vanishing contribution so that the total charge is non zero.  In the MV model,
color charge screening is usually argued to occur on a scale size of order the confinement size scale and is outside the
range of applicability of the model.

 On the other hand, if one evolves the MV model~\cite{Iancu:2000hn,Ferreiro:2001qy}, one restores charge neutrality on a
 scale of order the saturation momentum~\cite{Iancu:2002tr}.  The effect of evolution is to modify the charge-charge 
 correlation function, and is best understood in transverse momentum space:
 \begin{equation}
  \langle \rho^a(y_1,\vec{k}_{1}) \rho^b(y_2, \vec{k}_{2}) \rangle = (2\pi)^2 {{d\mu^2} \over {dy}} \delta^{ab} \delta(y_1-y_2)  \delta^{(2)} (\vec{k}_1+\vec{k}_2)  \Delta(\vec{k}_1)\,.
 \end{equation}
 The total charge in a transverse plane corresponding to a fixed rapidity is the limit as $k \rightarrow 0$ of the Fourier
 transformed charge density.  The requirement of charge neutrality is that
 \begin{equation}
    \lim_{k \rightarrow 0} \Delta(k) \rightarrow 0\,.
 \end{equation}
 
It should be noted here that we are not being careful about gauge invariance when discussing a charge charge correlation
function at non-zero spatial separation.  This is fixed up by inserting line ordered phases attached to the charge densities. 
This complication comes in when one goes beyond leading order in coupling in the MV model; in this case fluctuations in the 
transverse vector potential become important.  

In a recent paper, we wrote down a phenomenological model which includes corrections to the MV
that enforce charge neutrality~\cite{McLerran:2015sva}.  This model has
\begin{eqnarray}
  \Delta(k) & = &  {k^2 \over {k^2+m^2}} \nonumber \\
  & = &  1 - {m^2 \over {k^2+m^2}} \,. 
\end{eqnarray}
Here $m$ is of order the saturation momentum.  The charge induced by the first term on the RHS of this equation is precisely
cancelled by the second term.  Note that in coordinate space, the second term Fourier transforms to a $K_0(mr)$ so that the 
neutralizing polarization charge density 
is exponentially well localized around the initial charge density. 

One effect of including the polarization charge density is to generate angular moments of elliptic flow
for radiation from the CGC~\cite{McLerran:2015sva}.  Without final state interactions, one generates even moments
for multiparticle correlations which at high momentum have the correct semiquantitative features to 
describe flow harmonics seen in $pp$ and $pA$ collisions.  However these correlations are only present if we further
generalize our model to a finite number of emission sources.  As happened for ellipticities in Glauber or CGC 
computations~\cite{Bzdak:2013rya,Yan:2014afa}, fluctuation induced ellipticities vanish for large numbers of emitters. 
It also should be noted that the odd flow moments vanish and may require some degree of final state interaction
and a deeper understanding of the nuclear wave function. 
In any case, whether these flow contributions are essentially modified by final state interactions or not, it is important
to understand the initial state contribution to such moments.

It is the purpose of this paper to properly motivate a theory of a finite number of color charge emitters.
The $m^2$ correction to the charge density correlation function generates an effective Coulomb interaction.   
This Coulomb interaction must be
viewed as arising from a higher order term in the operator product expansion for the   effective theory of sources
in the McLerran-Venugopalan model.  It does not directly arise from Coulomb interactions of the QCD Lagrangian from 
which this theory is derived.  The screening effect we generate from our
effective action  is an analogy to Debye  screening for a thermalized Quark Gluon Plasma.

\section{The Coulomb Interaction and Charge Screening}

We begin by observing that the action that generates the screened propagator above is 
\begin{equation}
	\label{eq:model}
	S  =  \int dy \int \frac{d^2k_\perp}{(2\pi)^2} \,\, 
	  \rho (y,\vec{k}_\perp) \left[  
	  {\frac{1}{2 \overline{\mu}^2} }
  \left( 1 + \kappa {  \overline{\mu}^2 \over k_\perp^2} \right) \right]
	  \rho (y,- \vec{k}_\perp)\, ,  
\end{equation}
where
\begin{equation}
  \overline{\mu}^2 = {{d\mu^2} \over {dy}}
\end{equation}
is the two dimensional charge density per unit rapidity. And we introduced the dimensionless coefficient 
$\kappa = m^2 /  \overline{\mu}^2 $. 
Note that the {\it second} term in this equation is simply the two-spatial dimensional Coulomb interaction which in coordinate space is
\begin{equation}
	S_{C} = \int dy\,  \int d^2x_\perp d^2y_\perp\, \,  
	\rho(y,\vec{x}_\perp) 
	\left[
		{\kappa \over {8\pi}}
		\ln\frac1 {(\vec {x}_\perp-\vec{y}_\perp)^2 \Lambda_{\rm IR}^2} 
	\right]
	\, \, \rho(y,\vec{y}_\perp)\,.
\end{equation}
The dependence upon the IR scale arises from regulating the integration over small $k_\perp$ in the Fourier transformed 
action integration.  Note that if it is much larger than all scales in the problem,
it projects the total color charge in a unit of rapidity onto zero.  Alternatively, if the total color charge in a 
unit of rapidity is zero, which we will assume in what follows, then there is no dependence upon the IR scale, which
may therefore be replaced by some convenient physical scale size of order the size of
the system being considered.

The parameter $\bar\mu^2 = d\mu^2/dy$ is proportional to the charge  squared  per unit area per unit rapidity, 
and is therefore proportional to the saturation momentum.   We can see this explicitly
for adjoint representation sources (see e.g.~\cite{Iancu:2002tr,McLerran:2015sva}),
where
\begin{equation}
   \mu^2 = 4\pi \alpha_s N_c N/S_\perp. 
 \end{equation}
 Here $N$ is the total number of particle integrated over rapidity, and $S_\perp$ is the transverse area.
 The saturation momentum in the MV model up to logarithmic corrections (see e.g.~\cite{Iancu:2002tr})  is
 \begin{equation}
   Q_s^2 = \alpha_s N_c \mu^2. 
 \end{equation}
% Since the multiplicity of gluons grows as $dN/dy \sim e^{\kappa^\prime \alpha_s y}$
%where $\kappa^\prime$ is a constant determined by evolution equations, we conclude that
 %up to a contant factor $\kappa$,
%\begin{equation}
%	S_{C} = -{ \kappa \over 2} \int~ dy~\int d^2x_\perp d^2y_\perp {1 \over {4\pi}}
%		 \ln\left[(\vec {x}_\perp-\vec{y}_\perp)^2 \Lambda_{\rm IR}^2\right]  \rho(y%,\vec{x}_\perp) \rho(y,\vec{y}_\perp)\,.
%\end{equation}

Note that for a truly two dimensional Coulomb interaction there is a dimensional scale associated
with the two dimensional coupling.  This has disappeared from this formula, and has been replaced by the
dimensionless three spatial dimensional coupling.  We can think of our effective action as being associated
with a dimensional scale of order the saturation momentum, which is then weighed
in the action like $e^{-E/T}$ where the effective temperature $T$ is proportional to the saturation momentum.

The Coulomb 2-dimensional interaction is natural to expect when we have Lorentz
 boosted Coulomb fields.  The electric field around a point charge is boosted to 
\begin{equation}
   F^{i+} \propto \delta(x^-)    {\hat{r} \over r}
\end{equation}
and arises from  a vector potential $A^+ \propto \ln(r)$.  This vector potential has an infrared cutoff
originating from a vector potential of the form
\begin{equation}
	A^+ \propto \frac{\gamma} {\sqrt{\gamma^2x^{-2} + x_T^2}} %-1 /x^-
\end{equation}
In the limit $\gamma \rightarrow \infty$, this reduces to the logarithmic potential, so long
as $r \gg x^-/\gamma$.  This explains the infrared cutoff in the two dimensional potential.
The origin of locality in rapidity follows because $x^- \sim 1/\gamma$ implies that
$\Delta x^- = \tau_1 e^y_1 - \tau_2 e^y_2 \sim 1/\gamma$ or $\Delta y = \ln(\tau_1/\tau_2)$.
This means interactions are localized within about a unit of rapidity.

Although it might appear that the Coulomb interaction 
under investigation might originate from a simple  interpretation of the underlying gauge theory,
it is not the case, because $S_C$ is not  of the covariant form $F^{i+} F^{i-}$.  It should be rather 
thought of as arising from an operator product expansion for the effective action of the sources.

\section{Lattice formulation for  the effective action and numerical results}

This theory can be put on a finite grid.  There is a mild logarithmic singularity at short distance in 
the Coulomb potential  can be regulated by the finite grid size in the transverse direction.  We can 
choose the spacing to be one unit of rapidity, and this is a good approximation since distributions
are slowly varying over one unit of rapidity, and the Coulomb interaction, for the reasons discussed 
above should have a range of about one unit of rapidity if the interaction in rapidity is properly accounted for. 
This is of course finite in the limit the grid size shrinks to zero.  We also assume the IR divergence is taken
care of by requiring total color charge neutrality on a confining scale.

It is useful to think about this theory for a discrete number of emitters, that is we imagine that the sources
of the color charge are single gluons carrying an amount of charge corresponding to the octet
representation of the gauge group.  This pushes the classical description of the color charge 
to its limit of validity.  To have a truly proper theory one should treat the color charge 
operator as a matrix and solve the theory like was done in static matrix models.  For our 
proposes, we will however treat the color charge as a continuous classical variable.  Our
interest is in understanding the effects of having a finite number of particles and writing 
down a theory that corresponds to such a finite number of sources.  Presumably, if we discretize 
on a small scale, but consider quntlities whose scale of variation
is over a size scale that includes many units of color charge, treating the color sources as classical
variable should be well behaved.

Before the discretization is performed, it is necessary to regularize the interaction 
and maintain positivity of the action. The last requirement does not allow one to use a naive
delta-function point-like source on the lattice. 
We can maintain the positivity of the action   when we spread the sources 
over the region of size $\Lambda^{-1}_{\rm UV}$, i.e. in the momentum space we consider 
\begin{equation}
\label{Eq:rho_cut}
\rho^a(y, {\vec k}_\perp) = 
\sum_{i} \frac{\Lambda^2_{\rm UV}}{k^2 + \Lambda^2_{\rm UV}}
\xi_i^a(y) \exp(-i {\vec k}_\perp {\vec x}_i)  
\end{equation}
or transforming into the coordinate space  
\begin{equation}
\label{Eq:rho_cut_x}
\rho^a(y, {\vec x}_\perp) = 
\frac{1}{2\pi} \sum_{i} \xi_i^a(y) K_0(|{\vec x}_\perp - {\vec x}_i|) \Lambda_{\rm UV}\,.   
\end{equation}
Note that in the limit of large $\Lambda_{\rm UV}$ Eq.~\eqref{Eq:rho_cut} reduces to the Fourier transform
of the delta-function.

Thus, combining Eq.~\eqref{Eq:rho_cut_x} and Eq.~\eqref{eq:model}
results in the MV action which is manifestly positive 
\begin{equation} 
\label{eq:S2int}
S_{\rm MV}[\{ \xi\}, \{ { \vec x } \}] = \int dy  \; \sum_{i,j} 
\xi_i^a(y)\, \, \left[ 
%\frac1{8 \pi} 
\frac{
|\hat{\vec x}_i - \hat{\vec x}_j|
K_1(|\hat{\vec x}_i - \hat{\vec x}_j|)
}{ 8 \pi  \hat{\mu}^2}
\right]
\, \, 
\xi_j^a(y),
\end{equation}
where in order to simplify the notations  we introduced the following dimensionless variables 
$\hat{x} = x \Lambda_{\rm UV}$ and $\hat{\mu} = \bar{\mu}/\Lambda_{\rm UV}$.
The Coulomb term is then 
\begin{eqnarray}
	&&S_C [\{ \xi\}, \{ { \vec x } \}] = 
	 \int dy \int d^2 x_\perp d^2 y_\perp\,\, \rho(y, {\vec x}_\perp) 
\left[
	 \frac{\kappa}{4\pi}
 \ln \frac{1}{|{\vec x}_\perp-{\vec y}_\perp|\Lambda_{\rm IR}} \right]
\rho(x_-, {\vec y}_\perp) = 
\\ &&
\int dy
\sum_{i,j} 
\xi_i^a(y)
%\\ 
%&&
\left[
	 \frac{\kappa}{4\pi}
\left( 
\ln \frac{1}{|\hat{\vec x}_i-\hat{\vec x}_j|\hat\Lambda_{\rm IR}}
- 
K_0 \left( |\hat{\vec x}_i-\hat{\vec x}_j| \right)
- \frac12 |\hat{\vec x}_i-\hat{\vec x}_j| 
K_1 \left( |\hat{\vec x}_i-\hat{\vec x}_j| \right)
\right)
\right] \xi_j^a(y)\,. \nonumber 
	\label{eq:coul}
\end{eqnarray}

We note that the kernel enclosed in the rectangular brackets is finite at small distances 
due to compensation of the logarithmic singularities in the first and second terms. We will return to this point in 
the next section.

Now to perform the simulations, one can use  a spatial rectangular lattice with the lattice spacing $a$ and 
transverse length $L$.
The rapidity is to be discretized as well to simulate a finite extent of sources  in the rapidity direction. 
This means that the integrals with respect to rapidity are replaced by sums as follows 
\begin{equation}
	\int dy \to \sum_n \delta y\, . 
	\label{eq:intdy}
\end{equation}

It is interesting that the action for a finite number of sources is similar to that for the MV model except that in the MV model, sources are allowed to fluctuate in all of the two dimensional space.  The 
typical charge per grid unit in the MV model is the same as the typical charge per particle in our discretized model.  The essential difference is that for our discretized model, only a finite fraction of the transverse area is covered by sources.  This is what ultimately will allow for generation of
elliptical flow moments in radiation from the grid, for effect that vanish in the limit that the number of sources  $N \rightarrow \infty$.

%\section{Numerically Computing the S Matrix}

As a test of our formalism, we will compute the source-source correlator and 
the S matrix for single particle scattering by numerically
solving the finite $N$ discretized version of our theory. 

Consider $N$ sources distributed in the target. We will assume that the distribution is uniform in rapidity and thus 
the number of sources per a rapidity slice is given by $N/N_y$ where $N_y$ is the number of rapidity slices. 

The sources are spread as we discussed before. Their (non-Abelian) Weizs\"acker-Williams fields are %pure gauges;
in covariant gauge,
\begin{equation}\label{eq:A+} 
	A^{+ a}(y,\vec{x}_\perp) = - \frac{g}{ \nabla_\perp^2} \rho^a(y,\vec{x}_\perp)~.
\end{equation}
The only non-vanishing field
strength is $F^{+i}=-\partial^i A^+$. The (light-cone) electric field
is
\begin{equation}
E^i = \int d x^- F^{+i} = - \partial^i \int d x^- A^+~.
\end{equation}
The propagation of a fast charge in this field is described by an
eikonal phase given by a light-like Wilson line $V(\vec{x}_\perp)$:
\begin{equation}\label{eq:V_rho}
	V(\vec{x}_\perp) = \mathbb{P} \exp\left\{ - ig \int d y  
		A(y,\vec{x}_\perp) \right\} \,,
\end{equation}
where $\mathbb{P}$ denotes path-ordering in rapidity $y$ and the field can be found by solving Eq.~\eqref{eq:A+} 
\begin{equation}
	A(y,\vec{x}_\perp) = \frac{g}{2\pi} \sum_i \xi_i(y)  
	\left( \ln\left[\frac{1}{|\hat{\vec x}_\perp-\hat{\vec x}_i| \hat{\Lambda}_{\rm IR} } \right] 
		- K_0(|\hat{\vec x}_\perp-\hat{\vec x}_i|)\right)\,.
	\label{Axminus}
\end{equation}
Here $\hat{\Lambda}_{\rm IR} = \Lambda_{\rm IR}/\Lambda_{UV}$.
Note that 	$A(y,\vec{x}_\perp)$ is finite at any $\vec{x}_\perp$ except for infinite distances  owing to the following identity for small ${\hat {r}}$
\begin{equation}
	\ln\left[\frac{1}{\hat{r} \hat{\Lambda}_{\rm IR} } \right] 
 - K_0(\hat{r}) = 
		\gamma+\log \left(\frac{1}{2 \hat{\Lambda}_{\rm IR} }\right)+
		\frac{1}{4}\hat{r}^2 \left(\log
		\left(\frac{\hat{r} }{2 e }\right)+\gamma\right)+{\cal{O}}(x^3). 
	\label{eq:K0expansion}
\end{equation}

The S-matrix for scattering of this
charge off the given target field configuration is
\begin{equation}
	{\cal S}_\rho(\vec{r}_\perp,\vec{b}_\perp) \equiv \frac{1}{d_R}\, {\rm tr} \, V^\dagger(\vec{x}_\perp)\,
	V(\vec{y}_\perp)~, ~~~~\vec{r}_\perp \equiv \vec{x}_\perp-\vec{y}_\perp~,~~~2\vec{b}_\perp \equiv \vec{x}_\perp+\vec{y}_\perp~.
\end{equation}
In momentum space, 
this can be written in the following form 
\begin{equation}
	{\cal S}_\rho(\vec{k}_\perp)  =  \frac{1}{d_R}\, {\rm tr} \left[ \int d^2x_\perp d^2y_\perp e^{i \vec{k}_\perp (\vec{x}_\perp-\vec{y}_\perp)}   V^\dagger(\vec{x}_\perp)\,
	V(\vec{y}_\perp)\right] = \frac{1}{d_R}\,  {\rm tr} \,   V^\dagger(\vec{k}_\perp)\,
	V(\vec{k}_\perp), 
\end{equation}

The S-matrix is obtained by averaging ${\cal S}_\rho(\vec{k}_\perp)$ with respect to target configurations. 
Those are to be generated using the effective action we defined above. Here it is 
important to realize that the effective action is quadratic in color charges $\xi_i$,
and thus they can be integrated out, leaving us with a non-trivial complicated function of  
the positions of the charges. We prefer not to follow this way but rather to simulate the configurations 
using the Metropolis Monte-Carlo algorithm as detailed in Appendix. 

\begin{figure}
	\centerline{\includegraphics[width=0.5\linewidth]{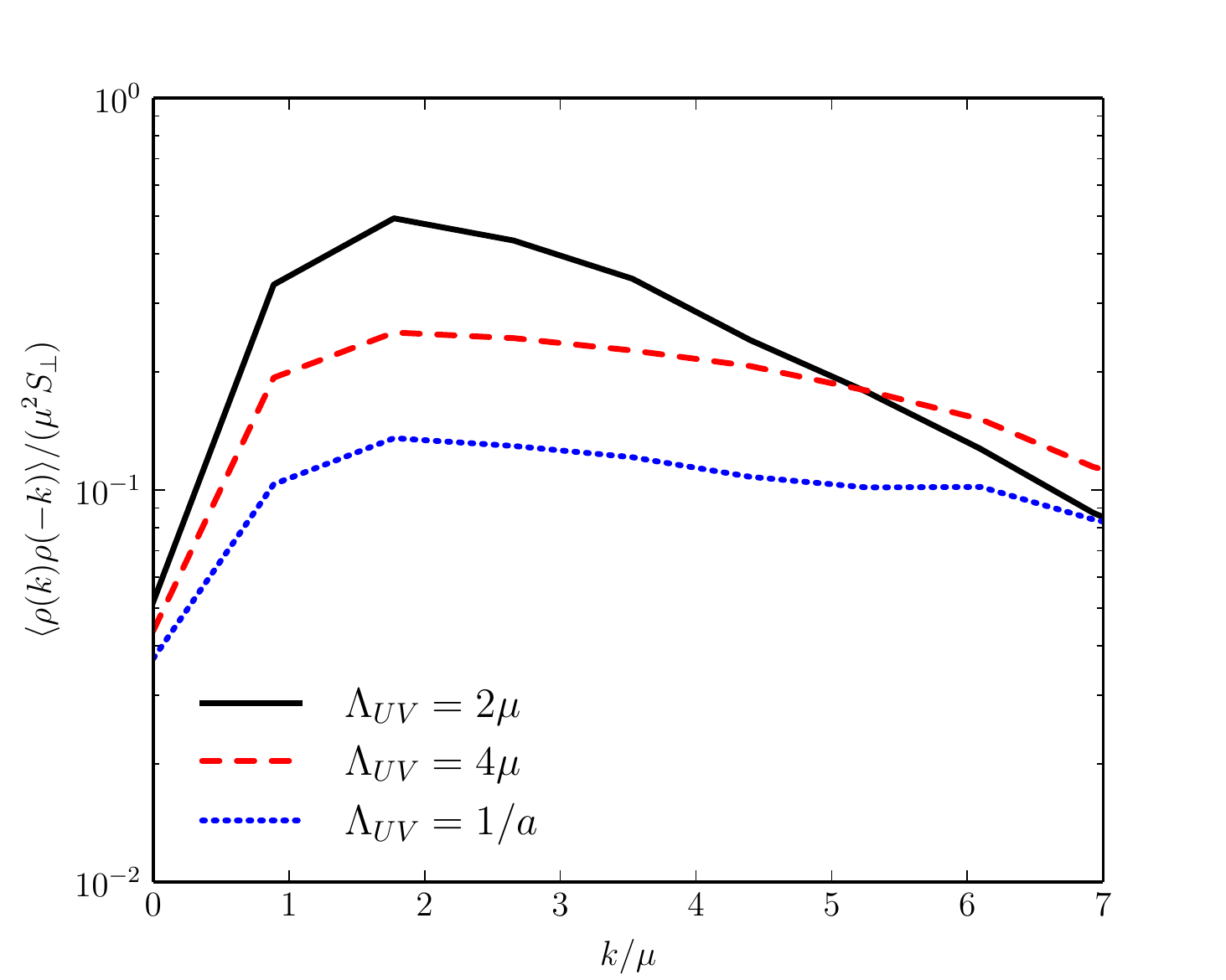}}
\caption{The correlator $\langle \rho(\vec{k}) \rho(-\vec{k}) \rangle$ normalized to yield $1$ in the limit of the infinite number 
of sources. The number of sources is $N=500$ and $\kappa=1$.     }
\label{fig:lambda_uv_dep}
\end{figure}

Lets first consider the source-source correlator for different number of particles $n$ per rapidity slice.
We check that there is a convergence if large values of $\Lambda_{\rm UV}$ are considered. 
From very general considerations, it is clear  that if the momentum $k_\perp$ is larger than 
$\Lambda_{\rm UV}$ model results will be sensitive to the details of the source regularization
in Eq.~\eqref{Eq:rho_cut}. Indeed Fig.~\ref{fig:lambda_uv_dep} for $\Lambda_{\rm UV}=2\mu$ 
shows a very significant variation in 
the correlator $\langle \rho(\vec{k}) \rho(-\vec{k}) \rangle $ which, at large $k_\perp$,  is expected to be independent 
of the momentum
and  assume $\mu^2 S_\perp$ in the limit of the infinite number of sources. Nonetheless independent of the scale 
$\Lambda_{\rm UV}$ we see the color neutralization driving the value of the correlator to zero at small momentum.  

In our calculations we prefer to set the scale  $\Lambda_{\rm UV}$ to the largest possible on the lattice $1/a$. In this case,
the $\rho \rho$ correlator is approximately constant above the color neutralization scale.  

\begin{figure}
	\centerline{
		\includegraphics[width=0.5\linewidth]{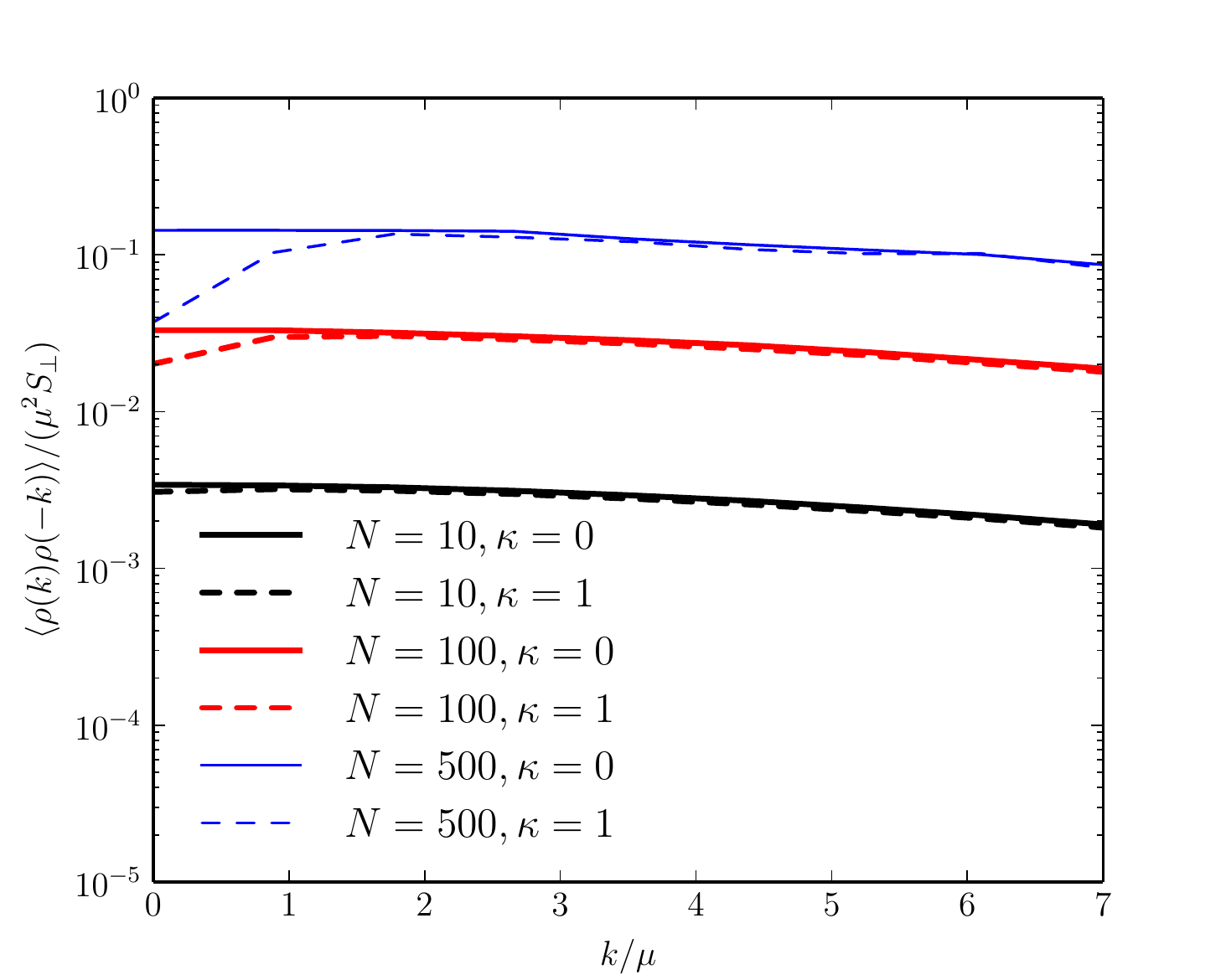}}
\caption{The correlator $\langle \rho(\vec{k}) \rho(-\vec{k}) \rangle$ normalized to yield $1$ in the limit of the infinite number 
of sources. The solid (dashed) lines correspond to MV model without (with) Coulomb interaction. 
The infrared cutoff  is $\Lambda_{\rm IR} = 1/L$, 
where $L$ is the system size.  
}
\label{fig:rhorho}
\end{figure}

The dependence of the correlator on the number of sources is displayed  in Fig.~\ref{fig:rhorho}
for $\kappa=0$ and $1$. 
The offset of the Debye screening can be seen for the number of particles $N=100$.  

\begin{figure}
	\centerline{
		\includegraphics[width=0.5\linewidth]{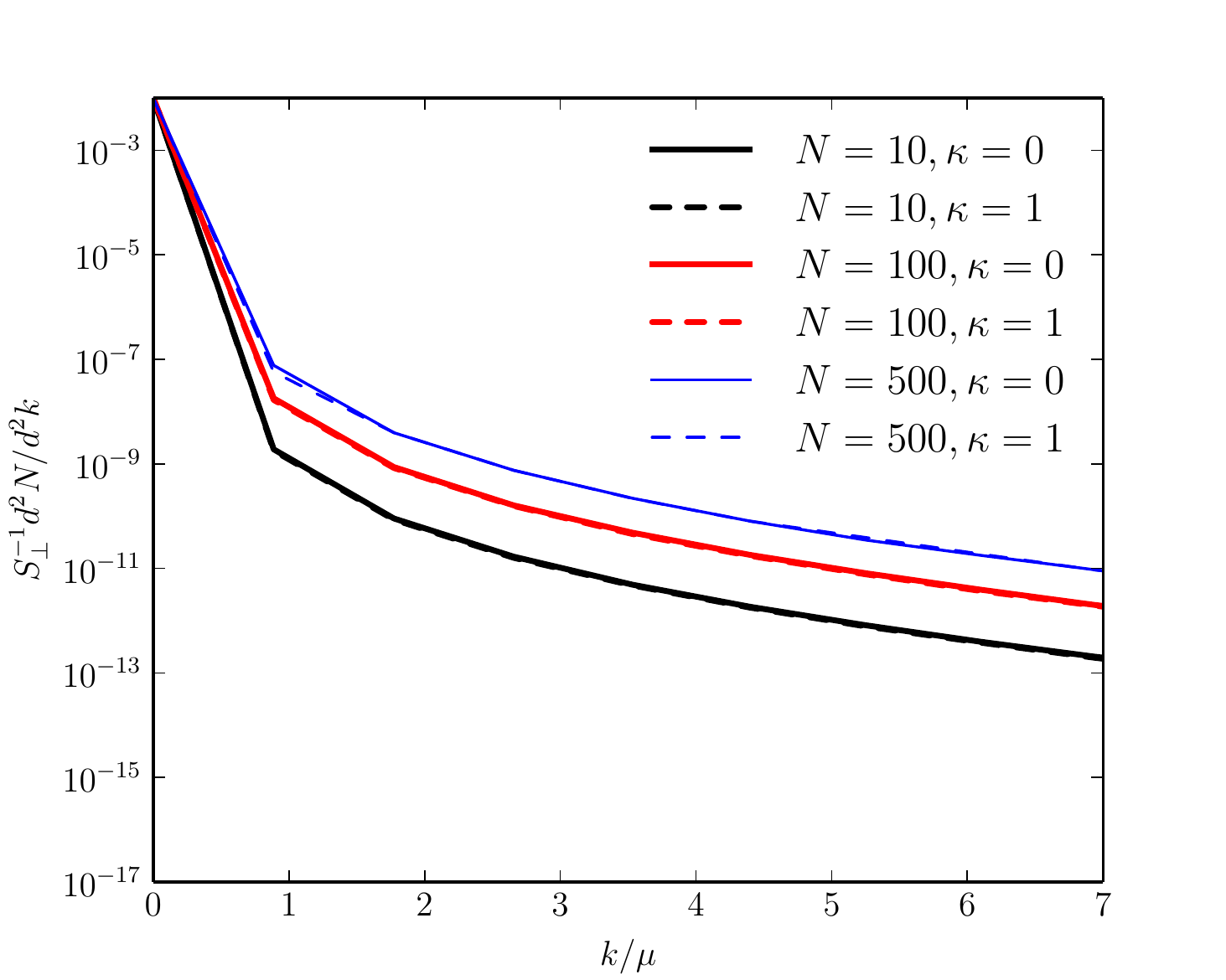}}
\caption{ The differential spectrum $d^2 N/d^2 k$  normalized by the transverse area of the target.
The solid (dashed) lines corresponds to MV model without (with) Coulomb interaction. 
The ultraviolet and infrared cutoffs are set to $\Lambda_{\rm UV} = 4 \mu$ and $\Lambda_{\rm IR} = 1/L$, 
where $L$ is the system size.  
}
\label{fig:S}
\end{figure}

To check that the model reproduces $1/k_\perp^4$ dependence of the particle spectrum, we 
computed the S matrix, depicted in Fig.~\ref{fig:S}. As expected, the results are not very sensitive to the 
Coulomb interaction. This  however should not be discouraging since the effect Coulomb interaction is essential
for the two particle correlation function, as discussed in Ref.~\cite{McLerran:2015sva}. 

\section{Conclusions}
We considered an extension of the McLerran-Venugopalan to include a finite number of sources of color charge and 
Coulomb interactions between  color charges. To simulate this theory on the lattice, the sources are regularized 
 in a manner that maintains the positivity of the MV action. We showed that if the corresponding scale of the transverse momentum is 
large it leads to the expected results of approximately constant $\rho\rho$-correlator. We investigated the dependence on 
$\kappa$ and showed the emergence of the color neutralization in the model. In this paper, we did not consider  
the two particle correlation function, this is the ultimate aim of the model which will be reported elsewhere. 

\section{Acknowledgements}

The authors gratefully acknowledge the comments of Jean Paul Blaizot, Soren Schlichting, Bjoern Schenke, Raju Venugopalan and Ulrich Heinz.

The authors are supported under Department of Energy contract number Contract No. DE-SC0012704.

\section{Appendix A: Numerical procedure}

The averaging with respect to the target configurations 
must be performed with the weight function 
\begin{equation}
	W[\{x\},\{\xi\}] = \exp(-S(\{x\},\{\xi\})). 
	\label{eq:W}
\end{equation}
In order to generate the configurations we use the Metropolis algorithm for each rapidity slice. 
Although the functional $S(\{x\},\{\xi\})$ is quadratic in $\{\xi\}$, it has a very non-trivial structure in the sources' positions $\{x\}$.
The Metropolis algorithm is performed along the following steps:
\begin{enumerate}
	\item Initially $N/N_y$ sources are randomly distributed on a two-dimensional 
		rectangular grid with the spatial extent $L$ and the spacing $a$. $N_y$ is the number of the rapidity slices, see below.  For each point, 
		the initial color vector $\xi^a$ is generated according to the Gaussian  probability distribution  
		with zero mean and the variance $\bar{\mu}$.  
	\item The Metropolis step is performed. One source is randomly chosen; 
		a trial change, $\Delta S$,  of the action $S = S_{\rm MV} + S_C$  is computed; if the source position is 
		randomly adjusted by a step  $a$ in a randomly chosen direction and its color vector 
		receive a random contribution distributed according to a Gaussian with zero mean and 
		the variance of 10\% of $\bar{\mu}$.  
	\item If $\Delta S < 0$, the change of the position and the color vector is accepted. 
		Otherwise the change is accepted with the probability $\exp(-S)$. 
	\item The steps 2 and 3 are repeated until equilibrium is reached. 
	\item The configurations are collected. To avoid autocorrelation, 
		there is at least 10 MC sweeps between each saved configuration. 
\end{enumerate}
After the configurations are collected, we compute the source density $\rho(y, \vec{x}_\perp)$ and the corresponding $A(y, \vec{x}_\perp)$. 
The number of the rapidity slices is finite in our approach and each rapidity slice has approximately one unit in rapidity $\delta y = 1$.
We assume that the sources are generated in each rapidity slice independently. 
Overall we consider $N_y$ rapidity slices, we fix $N_y=5$. Owing to the path ordering the Wilson line is given by the product of 
exponentials computed in each rapidity slice 
\begin{equation}
	V(\vec{x}_\perp) = \prod_{i=1}^{N_y}  \exp\left(  
	- i g A(y_i,\vec{x}_\perp)   \right), \, \, y_i = i \delta y.   
	\label{eq:Vy}
\end{equation}
By performing Fourier transformation and averaging with respect to the target configurations we can extract $S(\vec{k}_\perp)$.


\begin{thebibliography} {000}

%\cite{McLerran:1993ni}
\bibitem{McLerran:1993ni}
  L.~D.~McLerran and R.~Venugopalan,
  %``Computing quark and gluon distribution functions for very large nuclei,''
  Phys.\ Rev.\ D {\bf 49} (1994) 2233
  doi:10.1103/PhysRevD.49.2233
  [hep-ph/9309289].
  %%CITATION = doi:10.1103/PhysRevD.49.2233;%%
  %1471 citations counted in INSPIRE as of 04 Dec 2015
  
  %\cite{McLerran:1993ka}
\bibitem{McLerran:1993ka}
  L.~D.~McLerran and R.~Venugopalan,
  %``Gluon distribution functions for very large nuclei at small transverse momentum,''
  Phys.\ Rev.\ D {\bf 49} (1994) 3352
  doi:10.1103/PhysRevD.49.3352
  [hep-ph/9311205].
  %%CITATION = doi:10.1103/PhysRevD.49.3352;%%
  %1086 citations counted in INSPIRE as of 04 Dec 2015
  
 % \cite{Iancu:2000hn}
\bibitem{Iancu:2000hn}
  E.~Iancu, A.~Leonidov and L.~D.~McLerran,
  %``Nonlinear gluon evolution in the color glass condensate. 1.,''
  Nucl.\ Phys.\ A {\bf 692} (2001) 583
  doi:10.1016/S0375-9474(01)00642-X
  [hep-ph/0011241].
  %%CITATION = doi:10.1016/S0375-9474(01)00642-X;%%
  %821 citations counted in INSPIRE as of 04 Dec 2015
  
  %\cite{Ferreiro:2001qy}
\bibitem{Ferreiro:2001qy}
  E.~Ferreiro, E.~Iancu, A.~Leonidov and L.~McLerran,
  %``Nonlinear gluon evolution in the color glass condensate. 2.,''
  Nucl.\ Phys.\ A {\bf 703} (2002) 489
  doi:10.1016/S0375-9474(01)01329-X
  [hep-ph/0109115].
  %%CITATION = doi:10.1016/S0375-9474(01)01329-X;%%
  
  %\cite{Iancu:2002tr}
\bibitem{Iancu:2002tr}
  E.~Iancu, K.~Itakura and L.~McLerran,
  %``Geometric scaling above the saturation scale,''
  Nucl.\ Phys.\ A {\bf 708} (2002) 327
  doi:10.1016/S0375-9474(02)01010-2
  [hep-ph/0203137].
  %%CITATION = doi:10.1016/S0375-9474(02)01010-2;%%
  %332 citations counted in INSPIRE as of 04 Dec 2015
  
  %\cite{McLerran:2015sva}
\bibitem{McLerran:2015sva}
  L.~McLerran and V.~Skokov,
  %``Finite Numbers of Sources, Particle Correlations and the Color Glass Condensate,''
  arXiv:1510.08072 [hep-ph].
  %%CITATION = ARXIV:1510.08072;%%
  
  %\cite{Bzdak:2013rya}
\bibitem{Bzdak:2013rya}
  A.~Bzdak, P.~Bozek and L.~McLerran,
  %``Fluctuation induced equality of multi-particle eccentricities for four or more particles,''
  Nucl.\ Phys.\ A {\bf 927} (2014) 15
  doi:10.1016/j.nuclphysa.2014.03.007
  [arXiv:1311.7325 [hep-ph]].
  %%CITATION = doi:10.1016/j.nuclphysa.2014.03.007;%%
  %32 citations counted in INSPIRE as of 14 Dec 2015
  
  %\cite{Yan:2014afa}
\bibitem{Yan:2014afa}
  L.~Yan, J.~Y.~Ollitrault and A.~M.~Poskanzer,
  %``Eccentricity distributions in nucleus-nucleus collisions,''
  Phys.\ Rev.\ C {\bf 90} (2014) 2,  024903
  doi:10.1103/PhysRevC.90.024903
  [arXiv:1405.6595 [nucl-th]].
  %%CITATION = doi:10.1103/PhysRevC.90.024903;%%
  %11 citations counted in INSPIRE as of 14 Dec 2015



\end{thebibliography}
\end{document}